\documentclass[12pt]{article}
\usepackage{graphicx}
\usepackage{amsfonts}
\usepackage{amssymb}
\newcommand{\nin}{\noindent}
\newcommand{\be}{\begin{equation}}
\newcommand{\ee}{\end{equation}}
\newcommand{\bea}{\begin{eqnarray}}
\newcommand{\eea}{\end{eqnarray}}
\newcommand{\br}{\hskip .25cm/\hskip -.25cm}

\newcommand{\nn}{\nonumber\\}

\newcommand{\ol}{\overline}

\begin{document}

~

\nin KCL-PH-TH/2011-{\bf 32}

\begin{center}

{\bf\Large{Lifshitz-type Quantum Field Theories\\
in Particle Physics}}

\vspace{1cm}

{\bf Jean Alexandre}\\
King's College London, Department of Physics, London WC2R 2LS, UK\\
jean.alexandre@kcl.ac.uk

\vspace{2cm}

{\bf Abstract}

\end{center}

\nin This introduction to Lifshitz-type field theories reviews some of its aspects in Particle Physics.
Attractive features of these models are described with different examples, as the improvement of 
graphs convergence, the introduction of
new renormalizable interactions, dynamical mass generation, asymptotic freedom, and other features 
related to more specific models. On the other hand,
problems with the expected emergence of Lorentz symmetry in the IR are discussed, related  
to the different effective light cones seen by different particles when they interact.

\eject

\tableofcontents

\eject

\section{Motivations}

Condensed Matter has often provided Particle Physics with essential concepts and tools, as Spontaneous Symmetry Breaking 
or Renormalization Group studies, because both areas deal with a large number of degrees of freedom. 
Indeed, the correlation length of a solid state system
increases drastically near a phase transition, making it necessary to consider a large number of degrees of freedom, 
as in Particle Physics, where creation and annihilation of particles implies an infinite number of degrees of freedom.
Lifshitz-type theories, which have been widely studied these last few years, is one of the examples where Condensed Matter gave new ideas to
Particle Physics, and this short review explains different features of these theories, which consist in a class of Lorentz-violating 
extension of the Standard-Model. 

A Lifshitz critical point \cite{Lifshitz} is defined as a point in the phase diagram, where three phases meet: a disordered phase,
a spatially homogeneous ordered phase and a spatially modular ordered phase (the spatially homogeneous ordered phase corresponds 
to a modular phase with vanishing momentum). Solid state systems exhibiting a Lifshitz critical point are characterized by 
anisotropic properties, and are described by two different correlation lengths, in different space directions. 
A clear introduction on the topic is given in \cite{condmat}, 
where Renormalization Group studies involving two different scaling laws are discussed.

In Particle Physics, anisotropy between spatial coordinates, or between time and space, breaks Lorentz invariance and therefore 
is apparently not welcome. 
Nevertheless, such anisotropies in space time can be seen as a way to study phenomenologically 
Quantum Gravity effects \cite{QuantGrav} for example. 
General studies of Lorentz symmetry violating theories can be found in \cite{Kostelecky1}, where Lorentz invariance is 
broken by non-trivial vacuum expectation values of vectors or tensors, and where effects as vacuum birefringence are discussed,
as a consequence of Lorentz symmetry violation. 
Also, one can find in \cite{Kostelecky2} tables for upper limits for Lorentz-violating coefficients.
Finally, the recent experiment OPERA \cite{Opera}, where superluminal neutrino were apparently observed, 
is obviously a strong motivation for Lifshitz-type theories, which allow superluminal propagation
of particles.

Most of the Lifshitz theories studied in Particle Physics involve 3+1 or 2+1 dimensions, keeping Galilean invariance
and therefore breaking isotropy between space and time. The models in \cite{LorentzLifshitz}, though, involve an additional 
space dimension, and the anisotropy is between this extra dimension and the Lorentz symmetric 3+1 world. This approach is
closer to the original Lifshitz theory in Condensed Matter, for systems at equilibrium, featuring an anisotropy between space dimensions,
but we chose to focus this review on 3+1 or 2+1-dimensional Physics with Galilean invariance,
and therefore with anisotropy between space and time. 

Lifshitz-type theories feature the following attractive properties, compared to Lorentz-symmetric theories:
\begin{itemize}
\item Improvement of the convergence of loop integrals: Time and space derivatives do not have the same mass dimension
in Lifshitz-type theories, such that the latter
contain higher order space derivatives, compared to the Standard Model, lowering the degree of divergence 
of loop integrals. 
It is known that Lorentz invariant higher order derivatives introduce ghosts excitations \cite{FradkinTseytlin}, 
and are not acceptable, whereas Lorentz violating higher-order {\it space} derivatives only  
are mathematically consistent; 
\item Extension of the class of renormalizable interactions: Operators in a Lifshitz-type Lagrangian 
have higher mass dimensions than in 
Lorentz symmetric theories, hence irrelevant bare coupling constants in the latter case 
(with negative mass dimension) can become marginal or relevant in the Lifshitz case;
\item Possibility of dynamical mass generation: Since coupling constants which appear in the Standard Model become
dimensionfull in the Lifshitz context, they provide a natural mass scale to the theory. 
If massless particles interact, it can thus be expected that mass is dynamically generated, provided that it does not 
contradict a symmetry, as gauge invariance for example. 
\end{itemize}
On the other hand, Lifshitz-type theories feature the following potential problem: 
they lead to different effective light 
cones seen by different particle species when they interact, since space and time derivatives are not dressed 
equally by quantum fluctuations.
This point will be discussed for different models described in this review.

Finally, Horava's Lifshitz-type theory of Gravity \cite{HoravaGravity} has generated a huge interest in the community dealing 
with Modified Gravity, but this topic is not discussed here, besides a remark in the conclusion, as it deserves a review on its own. 

Section \ref{structure} deals with free fields, and sets the basic properties.
Interactions are introduced in Section \ref{scalint} for scalars, where the problem of effective light cone is described.
Spontaneous symmetry breaking and asymptotic freedom are also discussed, and the Liouville effective
potential is exactly calculated.
Interacting fermions are discussed in Section \ref{fermions}, where dynamical mass generation is shown in a four-fermion model,
as well as in a Yukawa model. Asymptotic freedom and effective dispersion relations are also discussed.  
Finally, in Section \ref{gauge} gauge theories are shortly reviewed, as well as an alternative to Lifshitz-type theory,
which contains higher-order space derivatives in the gauge sector, but where space and time scale equally. \\

\section{Free Lifshitz theories}\label{structure}

A way to implement anisotropy between space and time is to define a quadratic fixed point of renormalization group transformations, which
is invariant under the scaling $\vec x\to b\vec x$ and $t\to b^z t$, where $z$ is the critical exponent. In this context, 
the mass dimensions of coordinates are $[t]=-z$, $[\vec x]=-1$, and the case $z=1$ corresponds to 
a Lorentz symmetric theory.\\
In what follows, a dot represents a time derivative, $\Delta=\vec\partial\cdot\vec\partial$ is the Laplacian,
and the space time metric signature is $(1, -1, \cdots, -1)$. 

\subsection{Scalars}

The most general action for a free scalar field is
\be\label{scal}
S_{scal}=\frac{1}{2}\int dt d\vec x \left( (\dot\phi)^2-\phi\left[ \sum_{i=1}^{i=z}
\Lambda_i^{2(z-i)}(-\Delta)^i\right] \phi-m_s^{2z}\phi^2\right)~,
\ee
where the mass dimensions are $[m_s]=[\Lambda_i]=1$ and $[\phi]=(d-z)/2$. One can already note that the scalar field is dimensionless
in the situation where $d=z$, allowing then any function of $\phi$ as a scalar potential. We will come back to this 
point in subsection \ref{LL}. The propagator for the scalar field is then
\be
G_{scal}(\omega,\vec p)=\frac{i}{\omega^2-\sum_{i=1}^{i=z}\Lambda_i^{2(z-i)}p^{2i}-(m_s)^{2z}+i\epsilon}~,
\ee
where $\omega$ is the frequency and $\vec p$ the momentum, and the corresponding dispersion relation is 
\be
\omega^2=m_s^{2z}+\sum_{i=1}^{i=z}\Lambda_i^{2(z-i)}p^{2i}~.
\ee
If $\Lambda_1\ne0$, one can rescale the frequency as $\omega=\Lambda_1^{z-1}\tilde\omega$, which leads to
\be\label{moddisprel}
\tilde\omega^2=\mu_s^2+p^2+\frac{\Lambda_2^{2(z-2)}}{\Lambda_1^{2(z-1)}}p^4+\cdots~,
\ee
where $\mu_s=m_s^z\Lambda_1^{1-z}$ and the dots represent higher orders in the momentum.
The dispersion relation (\ref{moddisprel}) can be seen as a modification of Lorentz symmetric dynamics, with higher-order powers in the momentum
that could be a result of Quantum Gravity effects \cite{QuantGrav}, and which are negligible in the IR limit where
$p^2<<\Lambda_1^2$.

\subsection{Fermions}\label{freeferm}

We present here the free fermionic Lifshitz action for odd values of $z$, since for even values of $z$, the Dirac structure 
of the spinor does not appear in the space derivatives. We consider here only the case $z=3$, and define 
\be\label{ferm}
S_{ferm}=\int dt d\vec x \left( \ol\psi i\gamma_0\dot\psi-\ol\psi(\Lambda^2-\Delta)(i\vec\partial\cdot\vec\gamma)\psi
-m_f^3\ol\psi\psi\right)~,
\ee
where the mass dimensions are $[m_f]=[\Lambda]=1$ and $[\psi]=d/2$ (same mass dimension as in a Lorentz symmetric theory). 
The fermion propagator is
\be
G_{ferm}(\omega,\vec p)=i~\frac{\omega\gamma^0+(\vec p\cdot\vec\gamma)(\Lambda^2+p^2)+m_f^3}
{\omega^2-p^2(\Lambda^2+p^2)^2-m_f^6+i\epsilon}~,
\ee
and the dispersion relation 
\be
\omega^2=m_f^6+\Lambda^4p^2+2\Lambda^2p^4+p^6~.
\ee
Once again, in the situation where $\Lambda\ne0$, it is possible to recover approximate Lorentz invariant dynamics, since  
the rescaling $\omega=\Lambda^2\tilde\omega$ leads to
\be
\tilde\omega^2=\tilde m_f^2+p^2+\frac{2p^4}{\Lambda^2}+\frac{p^6}{\Lambda^4}~,
\ee
where $\tilde m_f=m_f^3\Lambda^{-2}$. This dispersion relation implies a superluminal propagation, since the 
product of phase and group velocities is
\be
v^2\equiv\frac{\tilde\omega}{p}\frac{d\tilde\omega}{dp}=1+\frac{4p^2}{\Lambda^2}+\frac{3p^4}{\Lambda^4}~>1~,
\ee
and the superluminal effects are energy dependent.

\subsection{Abelian gauge field}

For the sake of clarity, we consider here the situation where $d=z=3$.
A simple action for the free Abelian gauge field is
\be\label{Abel}
S_{Abel}=-\frac{1}{4}\int dtd\vec x \left(\frac{2}{e^2}F^{0k}F_{0k}+\frac{1}{g^2}\Delta F^{kl}\Delta F_{kl} \right)~, 
\ee
where $F_{\mu\nu}=\partial_\mu A_\nu-\partial_\nu A_\mu$ and the mass dimensions are $[A_0]=2$ and $[A_k]=0$.
The action (\ref{Abel}) is invariant under usual gauge transformations, and the dispersion relation is
$g^2\omega^2=e^2p^6$. The parameters $e$ and $g$ are in principle independent,
and since the gauge field is free, they do not get dressed. One can chose to rescale time and $A_0$ as
\be
t\to\frac{g}{e}~t~~~~\mbox{and}~~~~A_0\to\frac{e}{g}~A_0~,
\ee 
such that
\be
S_{Abel}\to-\frac{1}{4eg}\int dtd\vec x \left(2F^{0k}F_{0k}+\Delta F^{kl}\Delta F_{kl} \right)~.
\ee
If we then set $eg=1$, the gauge propagator in the Coulomb gauge is found by writing the appropriate tensor structure:
\be
D_{00}=A~~~~D_{0k}=Bp_k~~~~D_{ij}=Cp_kp_l+E\eta_{kl}~,
\ee
and solve the equations 
\be
D_{\mu\rho}~\frac{\delta^2S_{Abel}}{\delta A_\rho\delta A_\nu}=i\eta_{\mu\nu}~,
\ee
in order to find the functions $A,B,C,E$.
One obtains then, in the Coulomb gauge, which is appropriate for a Lorentz violating theory,
\bea
D_{00}&=&\frac{-i}{p^2}\left( 1-\frac{\omega^2}{p^6}\right) \\
D_{0k}&=&i\frac{\omega p_k}{p^8}\nn
D_{kl}&=&\frac{-i}{p^6-\omega^2}\left( \eta_{kl}+\frac{\omega^2p_kp_l}{p^8}\right)  \nonumber~.
\eea
If one wishes to consider Lorentz-violating corrections to the usual Maxwell Lagrangian, one can 
take into account several terms involving space derivatives only, as for example
\be
\tilde S_{Abel}=-\frac{1}{4}\int dtd\vec x \left(2F^{0k}F_{0k}+\Lambda^4_1F^{kl}F_{kl}
-\Lambda^2_2\partial_iF^{ik}\partial^jF_{jk}+\Delta F^{kl}\Delta F_{kl} \right)~, 
\ee
where $\Lambda_1,\Lambda_2$ have mass dimension 1. Once again, for $\Lambda_1\ne0$, one can recover a Lorentz-like dispersion
relation, with additional higher order powers of the momentum which are negligible when $p^2<<\Lambda_1^2$.

\section{Quantization of scalar theories}\label{scalint}

We start to study interactions in the framework of Lifshitz theories with scalar models. 
The proof for renormalizability of such theories, preserving locality and unitarity, can be found in \cite{AnselmiGeneral}

\subsection{Example}

We consider here renormalizable self-interactions, for $d=3$ and $z=2$, with the bare action 
\be\label{Sz=2}
S_{z=2}=\int dt d\vec x\left( \frac{1}{2}(\dot\phi)^2+\frac{M^2}{2}\phi\Delta\phi-\frac{1}{2}\phi\Delta^2\phi
-\frac{1}{2}m^4\phi^2-\sum_{n=3}^{10}\frac{g_n}{n!}\phi^n\right)~,
\ee
where the mass dimensions are $[g_n]=5-n/2$, $[M]=[m]=1$ and $[\phi]=1/2$.
Note that a more general action could also contain the renormalizable derivative interaction $\phi^5\Delta\phi$. 
We define the one-loop self energy as
\be
\Sigma=i\left( \delta m^4+z_0\omega^2+z_1p^2+\cdots\right) ~,
\ee
where dots represent higher orders in the the frequency and the momentum, and $[z_0]=0$, $[z_1]=2$.
At one-loop, only two graphs contribute to the self energy: $\Sigma=i(A+B)$, where
the graph $A$ is obtained with one propagator and one interaction $\phi^4$; the graph $B$ is obtained with two propagators and two
interactions $\phi^3$. We have, after a Wick rotation,
\bea
A&=&\frac{g_4}{2(2\pi)^4}\int \frac{d\omega d\vec p}{\omega^2+M^2p^2+p^4+m^4}\\
B(\nu,\vec k)&=&\frac{g_3^2}{2(2\pi)^4}\int \frac{d\omega d\vec p}{\omega^2+M^2p^2+p^4+m^4}\nn
&&~~~~~~~~\times\frac{1}{(\omega-\nu)^2+M^2(\vec p-\vec k)^2+(\vec p-\vec k)^4+m^4}\nonumber~,
\eea
where $\nu$ and $\vec k$ are the external frequency and momentum respectively.
The graph $B$ is finite, and therefore its contribution to the scalar mass can be neglected compared to the mass correction
provided by the graph $A$. The latter, which does not depend on the external frequency and momentum, 
can be calculated by integrating over the frequency $\omega$ first, to obtain
\be
A=\frac{g_4m}{8\pi^2}\int_0^{\Lambda/m}\frac{x^2dx}{\sqrt{1+\mu^2x^2+x^4}}\simeq \frac{g_4\Lambda}{8\pi^2}~,
\ee
where $\Lambda$ is the UV cut off and $\mu\equiv M/m$. 
Note that the mass correction diverges linearly and not quadratically, as a result of the
anisotropic model with $z=2$. A model with $z=3$ would give a logarithmic divergence, therefore improving divergences. 
General comments on this anisotropy considered as a regulator can be found in \cite{Visser}, and other comments on
smoothing of UV divergences are given in \cite{Chao}. \\
One-loop corrections to the derivative terms, which are finite, are obtained by isolating the $\nu^2$ and $k^2$ terms in an expansion 
of $B(\nu,\vec k)$, which gives, after an integration over the loop frequency $\omega$,
\bea
z_0&=&-\frac{g_3^2}{64\pi^2m^7}\int_0^\infty\frac{x^2dx}{(1+\mu^2x^2+x^4)^{5/2}}\nn
z_1&=&\frac{g_3^2}{192\pi^2m^5}\int_0^\infty x^2dx\frac{10x^6+\mu^2x^4+\mu^4x^2-3(3\mu^2+10x^2)}
{(1+\mu^2x^2+x^4)^{7/2}}~.
\eea
The dispersion relation obtained after one-loop dressing is therefore
\be
(1-z_0)\omega^2=m^4+\delta m^4+(M^2+z_1)p^2+\cdots
\ee 
One notes that, in order to find a Lorentz symmetric dispersion relation in the IR,
one needs $M^2+z_1>0$. Otherwise one would get, after rescaling of the frequency,
\be
\tilde\omega=\tilde m^2-p^2+\cdots
\ee 
and no IR Lorentz-like symmetry can be recovered.

\subsection{Effective light cones}\label{efflightcone}

We deal here with an essential issue about Lifshitz type theories, which is related to the existence of 
different effective light cones seen by different particles when these interact. This feature has been studied in 
\cite{IengoRussoSerone} for two interacting scalar fields, for $d=4$ and $z=2$, described by an action of the form
\bea
S_{1,2}&=&\int dt d\vec x\left( \frac{1}{2}(\dot\phi_1)^2-\frac{c_1^2}{2}(\vec\partial\phi_1)^2-\frac{a_1}{2}(\Delta\phi_1)^2
-\frac{g_1}{4}(\phi_1\vec\partial\phi_1)^2\right. \nn
&&~~~~~~~~+\frac{1}{2}(\dot\phi_2)^2-\frac{c_2^2}{2}(\vec\partial\phi_2)^2-\frac{a_2}{2}(\Delta\phi_2)^2
-\frac{g_2}{4}(\phi_2\vec\partial\phi_2)^2\nn
&&\left. ~~~~~~~~-\frac{1}{2}m_1^4\phi_1^2-\frac{1}{2}m_1^4\phi_2^2-V(\phi_1,\phi_2)\right)~,
\eea
where $[\phi_1]=[\phi_2]=1$, and renormalizable interactions are chosen in the potential $V(\phi_1,\phi_2)$.
The scale dependence of dressed couplings is then studied at one-loop, and the dressed speeds $\tilde c_1(t)$ and $\tilde c_2(t)$ 
run logarithmically with the energy scale, $t=\ln(\mu/\mu_0)$.
Ignoring quantum corrections to the time derivatives, the dispersion relations for the two scalars are of the form
\bea\label{c1c2}
\omega^2&=&\mu_1^2+\tilde c_1^2(t)p^2+\cdots\nn
\omega^2&=&\mu_2^2+\tilde c_2^2(t)p^2+\cdots~,
\eea
and it is not possible to rescale time, space or fields to set $\tilde c_1=\tilde c_2=1$, since the fields interact.
These speeds define effective light cones, as seen in the IR, and apparently play the role of the maximum speeds for the particle.
If one expects the IR values of these speeds to coincide, so that $|\tilde c_1-\tilde c_2|$ is consistent with experimental
bounds, it is shown in \cite{IengoRussoSerone}
that the bare couplings have to be fine tuned to extremely small values, which is not natural. 
As a consequence, it seems that Lorentz invariance is not recovered in the IR, although one should not forget that 
the dispersion relations (\ref{c1c2}) are valid in the IR only, and that increasing the energy 
of particles modifies these dispersion relations. Therefore it is not clear if the speeds $c_1$ and $c_2$ should actually 
play the role of maximum speed.\\ 
Studies of dispersion relations are made from bare Lifshitz-type Lagrangians in \cite{ChenHuang}, for scalar and
Yang Mills fields and general $z$, where the breaking of Lorentz invariance is related to
time delays in gamma-ray burst phenomenology.\\
Also, loop corrections arising from gravitons in Horava-Lifshitz Gravity are calculated in
\cite{Pospelov}, where the corresponding effects on the matter sector are studied. As a consequence of local Lorentz violation, it is
shown that different effective light cones are seen by different particles interacting with Horava-Lifshitz gravitons.

\subsection{Asymptotic freedom} 

A $CP^N$ model was studied in \cite{DasMurthy}, where asymptotic freedom and dynamical mass generation are proved to occur when $d=z$.
We will come back to dynamical mass generation when we study fermions, and asymptotic freedom is also shown in \cite{AnaFaraPasipTsap} 
for an $O(N)$ model in 2+1 dimensions and for $z=2$. We briefly review the latter results here.\\
Consider the following non-linear sigma model, in 2+1 dimensions and for $z=2$ \cite{AnaFaraPasipTsap}
\bea
S_{O(N)}&=&\int dtd\vec x\Big(\partial_0\vec e\cdot\partial_0\vec e-M^2\partial_i\vec e\cdot\partial_i\vec e
-\Delta\vec e\cdot\Delta\vec e\nn
&&~~~~~~~~+\eta_1(\partial_i\vec e\cdot\partial_i\vec e)^2
-\eta_2(\partial_i\vec e\cdot\partial_j\vec e)(\partial_i\vec e\cdot\partial_j\vec e)\Big)~,
\eea
where the constraint $\vec e\cdot\vec e=1$ is imposed. The mass dimensions are $[\vec e]=0$, $[M]=1$ and $[\eta_1]=[\eta_2]=0$.\\
A usual approach for these models is to introduce the fields $\sigma=e_0$ and the pions $\vec\pi=(ge_1,\cdots,ge_{N-1})$, 
where $g$ is the coupling constant of the model and the non-linear constraint reads $\sigma^2+g^2(\vec\pi)^2=1$. 
A perturbative expansion is then performed in \cite{AnaFaraPasipTsap}, for small $g$, and it is found that, for $\eta_1=1$ 
and $\eta_2=-2$, the beta function for $g$ is
\be
\beta(g)\equiv M\frac{\partial g}{\partial M}=-\frac{N-2}{4\pi}g^3+{\cal O}(g^5)~,
\ee
such that the model is asymptotically free, as the analogue 1+1 dimensional Lorentz symmetric model. Finally,
a current representation for the asymptotically free Lifshitz model can also be found in \cite{AnaFaraPasipTsap}.

\subsection{Dynamical symmetry breaking}

The one-loop effective potential in a scalar Lifshitz theory is calculated in \cite{EuneKimSon} for $z=2$, 
and it is found that quantum corrections generate a non-trivial minimum, 
as in the Coleman Weinberg mechanism \cite{Coleberg}. 
The article \cite{Faraxas} shows a similar calculation, for $z=2$ and $z=3$, and argue that in the case $z=2$
the symmetry is restored at finite temperature. \\
Starting from the massless bare potential $\lambda\phi^4/24$,
where $[\phi]=1/2$ and $[\lambda]=3$, the one-loop effective potential at zero temperature is \cite{Faraxas}
\be
V_0=\frac{\lambda}{24}\left( 1-\frac{15c}{2^5}\left( \frac{\lambda}{2\mu^3}\right)^{1/4}\right) \phi^4
-c\left( \frac{\lambda}{2}\phi^2\right)^{5/4}~,
\ee
where $c=10\Gamma^2(3/4)\pi^{-5/2}$ and the energy scale $\mu$ is used to define the renormalized coupling $\lambda$, 
$V^{(4)}(\sqrt\mu)\equiv\lambda$. This potential has a non-trivial minimum, and the scalar mass can be defined by expanding
the potential around the minimum.\\
The finite temperature one-loop effective potential is, in the high temperatures limit,
\be
V_T=V_0-\frac{\zeta(5/2)}{8\pi^{3/2}}T^{5/2}+\frac{2^{3/4}}{12\pi}T(\lambda\phi^2)^{3/4}~,
\ee
restoring the symmetric vacuum for $T$ large enough (the first temperature dependent term is a constant and does not play a role
for symmetry restoration). As remarked in \cite{Faraxas}, one must take care of non-analytical terms for $\phi=0$
in the one-loop effective potential, which cannot be taken into account to define the renormalized mass at $\phi=0$.
According to \cite{Faraxas}, no such symmetry breaking occurs for $z=3$, where also no symmetry restoration occurs at (one-loop)
finite temperature if there is symmetry breaking at the classical level.

\subsection{Exact effective potential}\label{LL}

We give here an example of a theory where the exact effective potential can be calculated, to all orders in $\hbar$.\\
As we have seen, the scalar field is dimensionless for $d=z=3$. For this reason, any differentiable function of 
the field is classically marginal, such that one can expect a renormalizable theory. 
We consider here the example of an exponential potential \cite{AlexFaraTsap}, as in the 
known Lorentz-symmetric Liouville theory in 1+1 dimensions \cite{Liouville}, 
where the scalar field is dimensionless.\\
The action we start with is
\be\label{SLL}
S_\mu=\int dtd\vec x
\left( \frac{1}{2}(\dot\phi)^2-\frac{1}{2}\partial^k\phi\Delta^2\partial_k\phi-\frac{\mu^6}{g^2} e^{g\phi}\right) ,
\ee
where $[\phi]=[g]=0$ and $[\mu]=1$, and an essential property of the model (\ref{SLL}) is the following:
a constant shift in the field $\phi(x)\to\phi(x)+\eta$ is equivalent to a redefinition of the
only dimensionful parameter as
\be
\label{mutilde}
\mu^6\to\tilde\mu^6=\mu^6e^{g\eta}~.
\ee
This identity, together with functional properties of the corresponding partition function, leads to \cite{AlexFaraTsap}
\be\label{Uprime}
U'=gU-\frac{g}{6}\dot U~,
\ee
where a prime denotes a derivative with respect to $\phi$ and a dot denotes a derivative with respect to the regulator
$t=\ln(2^{1/3}\Lambda/\mu)$.
The next step is to sum all the vacuum diagrams of the theory, to the leading order in $t$, 
which is possible since the only divergence in the theory is a one-loop graph made of one propagator only.
This resummation leads to the second essential identity \cite{AlexFaraTsap}
\be
\dot U=\frac{g}{8\pi^2}U'~[1+{\cal O}(t^{-1})]~.
\ee
Substituting this result in eq.(\ref{Uprime}) leads to a differential equation which integrates as
\be\label{Ufinal}
U=C(t)~ \exp(g_r\phi_0)~[1+{\cal O} (t^{-1})]~,
\ee
where $C(t)$ does not depend on $\phi_0$ and
\be\label{gr}
g_r\equiv g\left(1+\frac{g^2}{48\pi^2}\right)^{-1}~.
\ee
This relation between $g_r$ and $g$ is exact, and gives the field dependence of the renormalized potential.
In order to completely determine the latter, one still needs to specify the renormalized mass parameter $\mu_r^6$,
and the exact renormalized potential, to all loop orders, is finally
\be\label{Ur}
U_r=\frac{\mu_r^6}{g_r^2}\exp(g_r\phi)~.
\ee
One can check then \cite{AlexFaraTsap} that the one-loop theory agrees with these results. This example shows that 
the rich conformal structure in 1+1 dimensions is not necessary to derive the result (\ref{Ur}), which is obtained 
independently here.

\section{Interacting fermions}\label{fermions}

Lifshitz-type theories involving fermions exhibit two important features that we show here:
dynamical mass generation and asymptotic freedom. We will also find, as for the scalar case, the problem of effective light cone
for the Yukawa model.

\subsection{Four-fermion interaction}

We consider here the following action, describing a renormalizable four-fermion interaction, with $N$ flavours, 
in space dimension $d=3$ and for $z=3$, 
\be\label{free}
S_{4ferm}=\int dt d\vec x \left( \overline \psi_k i\gamma_0\dot\psi_k-\overline \psi_k(M^2-\Delta)(i\vec\partial\cdot\vec\gamma)\psi_k
+g(\ol\psi_k\psi_k)^2\right)~,
\ee
where $[g]=0$, $[M]=1$ and $[\psi]=3/2$. The indice $k$ runs from 1 to $N$. As we have seen in subsection \ref{freeferm},
this model implies superluminal propagation of fermions, and we note that the product of phase and group velocities
is not dressed at one-loop in this model, since the one-loop self energy does not depend on the external frequency and momentum.

\vspace{0.5cm}

\nin{\it Dynamical mass generation}\\ 
In order to study the possibility of generating a mass dynamically,
one can introduce a scalar auxiliary field $\phi$ to write the four-fermion interaction like
\be
\exp\left( ig\int (\ol\psi_k\psi_k)^2\right) =\int {\cal D}[\phi]\exp\left( i\int (2\phi\ol\psi_k\psi_k-g^{-1}\phi^2)\right)~,
\ee
where one notes that $\phi$ doesn't have the canonical mass dimension, since it doesn't have kinetic term, but rather has the dimension 
$[\phi]=3$. One integrates then over fermions to get the scalar effective potential
\be\label{Veff}
V_{eff}(\phi)=-g^{-1}\phi^2+\frac{N}{8\pi^3}\int d\omega ~p^2dp
\ln\left( \frac{\omega^2+(M^2+p^2)^2p^2+4\phi^2}{\omega^2+(M^2+p^2)^2p^2}\right) ~,
\ee
which must be regularized by a cut off $\Lambda$.
After taking the field derivative $V_{eff}'$ and integrating over the frequency $\omega$, 
one can see that the potential (\ref{Veff}) has a minimum $\phi_{min}$, which is solution of the gap equation
\be\label{gap}
\frac{2\pi^2}{Ng}=\int_0^\Lambda \frac{p^2dp}{\sqrt{(M^2+p^2)^2p^2+4\phi_{min}^2}}~.
\ee
This minimum, which is denoted $\phi_{min}\equiv m_{dyn}^3/2$, generates the 
dynamical mass term $2\phi\ol\psi_k\psi_k\to m_{dyn}^3\ol\psi_k\psi_k$, plus fluctuations\footnote{After integration of 
fermions, $\phi$ becomes dynamical and can fluctuate around $\phi_{min}$.}, which can be neglected 
for a large number of flavours $N$. Indeed, the problem depends on the effective coupling $\tilde g=Ng$, such that 
$g$ can be small if $N$ is large (for a fixed value of $\tilde g$), suppressing fluctuations.

\vspace{0.5cm}
 
\nin{\it Asymptotic freedom}\\
This property can be obtained from the interpretation of the gap equation (\ref{gap}) as a relation 
between the coupling $\tilde g$ and the cut off $\Lambda$,
for {\it fixed} dynamical mass $m_{dyn}$, since the latter is physical and should not depend on the cut off. 
Taking a derivative of the gap equation (\ref{gap}) with respect to $\Lambda$ gives then
\be
\beta(\tilde g)\equiv\Lambda\frac{\partial \tilde g}{\partial\Lambda}
=\frac{-\tilde g^2\Lambda^3}{2\pi^2\sqrt{(M^2+\Lambda^2)^2\Lambda^2+m_{dyn}^6}}~,
\ee
and the beta function is negative over all the range of energies $\Lambda$. 
For large values of $\Lambda$, we have
\be
\beta(\tilde g)\simeq-\frac{\tilde g^2}{2\pi^2}~.
\ee
One should bare in mind that this result is valid in the approximation where one can neglect fluctuations of the field $\phi$
around $\phi_{min}$, which can be justified with a large number of flavours.\\
Similar models have been studied in \cite{AnselmiMdyn,DarMandalWadia}, and
renormalizable four-fermion interactions are used in \cite{AnselmiNeutrinos} to generate dynamically neutrino masses. Also,
more general four-fermion interactions are described in \cite{AnselmiMdyn} and \cite{DarMandalNag}, 
where detailed discussions can be found.

\subsection{Yukawa model}\label{Yukawa}

We consider here a Lifshitz-type Yukawa action, in $d=3$ space dimensions and with anisotropic scaling $z=3$, and discuss
the main results derived in \cite{AlexFaraPasipTsap}. We start with the action
\bea\label{Y}
S_Y&=&\int dt d\vec x\Big\lbrace \frac{1}{2}(\dot\phi)^2-\frac{1}{2}(\vec\partial\phi)\cdot(\Delta^2\vec\partial)\phi
-\frac{1}{2}m_s^6\phi^2\nn
&&~~~~~~~~~+ i\ol\psi\gamma^0\dot\psi+i\ol\psi\Delta(\vec\gamma\cdot\vec\partial)\psi-g\phi\ol\psi\psi\Big\rbrace ~,
\eea
where the Yukawa coupling has mass dimension $[g]=3$, and $[m_s]=1$ and the fermion is massless. The main features of this model are the 
dynamical generation of a mass and of Lorentz-like kinetic terms for fermions.\\
Before studying non-perturbative aspects of this model, and
assuming for the moment that $m_{dyn}$ is indeed generated, one can remark that the model (\ref{Y}) is 
super-renormalizable, and that the only diverging graph is the one-loop scalar mass 
\bea
\tilde m_s^6&=&m_s^6-4g^2\int_{-\infty}^\infty
\frac{d\omega}{2\pi}\int \frac{d\vec p}{(2\pi)^3}
\frac{\omega^2+p^6-m_{dyn}^6}{\left(\omega^2+p^6+m_{dyn}^6\right)^2}\nn
&=&m_s^6-\frac{g^2}{\pi^2}\int_0^\Lambda \frac{p^8~dp}{[p^6+m_{dyn}^6]^{3/2}}\nn
&=&m_s^6-\frac{g^2}{\pi^2}\ln\left( \frac{\Lambda}{m_{dyn}}\right)+\cdots~,
\eea
where the dots are subdominant terms and $\Lambda$ is the UV cut off. As expected from this Lifshitz-type model, the scalar dressed mass 
does not diverge quadratically, but logarithmically only.

\vspace{0.5cm}

\nin{\it Dressed fermion sector}\\
We assume the following form for the effective action
\bea\label{SYeff}
S_Y^{eff}&=&\int dt d\vec x\Big\lbrace \frac{1}{2}(\dot\phi)^2-\frac{1}{2}(\vec\partial\phi)\cdot(\Delta^2\vec\partial)\phi
-\frac{1}{2}\tilde m_s^6\phi^2\nn
&&~~~~~~~~+ i\ol\psi\gamma^0\dot\psi-i\lambda\ol\psi(\vec\partial\cdot\vec\gamma)\psi+i\ol\psi\Delta(\vec\gamma\cdot\vec\partial)\psi\nn
&&~~~~~~~~-m_{dyn}^3\ol\psi\psi-\tilde g\phi\ol\psi\psi\Big\rbrace ~,
\eea
which takes into account the dressing of the scalar mass and the Yukawa coupling, but also 
the possibility for the generation of fermion dynamical mass $m_{dyn}$ and first order space derivative 
$i\lambda\ol\psi(\vec\partial\cdot\vec\gamma)\psi$, with $[\lambda]=3$. As a first approximation,
the effective action (\ref{SYeff}) does not take into account the dressing of 
time derivatives, or the scalar Lorentz-like kinetic term.\\
The generation of $m_{dyn}$ and $\lambda$ is studied with the Schwinger-Dyson equation, which is derived in \cite{AlexFaraPasipTsap},
\be\label{SD}
\Sigma_f=ig\tilde G_f\Theta\tilde G_s~,
\ee 
and gives an exact relation between the dressed fermion and scalar propagators $\tilde G_f,\tilde G_s$ 
and the dressed Yukawa coupling $\Theta$.  
$\Sigma_f$ is the fermion self energy, which, in the approximation (\ref{SYeff}), is  
\be
\Sigma_f(\vec k)=-\lambda(\vec k\cdot\vec\gamma)-m_{dyn}^3~.
\ee
In what follows, we neglect quantum corrections to the scalar propagator. Also, quantum corrections to the vertex
are neglected, which consists in the so-called Rainbow or Ladder approximation. The latter happens to be good in this case,
since one-loop corrections to the Yukawa coupling vanish \cite{AlexFaraPasipTsap}. Taking into account these assumptions, the
Schwinger-Dyson equation (\ref{SD}) leads, after integration over the loop frequency, to the two coupled self consistent 
relations \cite{AlexFaraPasipTsap}
\bea
\frac{4\pi^2}{g^2}&=&\int_0^\infty\frac{p^2dp}{A_sA_f(A_s+A_f)}\nn
\frac{4\pi^2}{g^2}\lambda&=&\int_0^\infty dp~p^8(p^2+\lambda)\frac{2A_s+A_f}{A_s^3A_f(A_s+A_f)^2}~,
\eea
where $A_s=\sqrt{p^6+\tilde m_s^6}$ and $A_f=\sqrt{p^2(p^2+\lambda)^2+m_{dyn}^6}$. A numerical study of these 
equations shows that $m_{dyn}$ and $\lambda$ are indeed generated above a critical value $g_c$ for the Yukawa coupling, which is 
of the order of $2\pi m_s^3$ \cite{AlexFaraPasipTsap}. The dispersion relation for the fermion is therefore
\be
\frac{\omega^2}{\lambda^2}=\left( \frac{p^2}{\lambda}+1\right)^2p^2+\frac{m_{dyn}^6}{\lambda^2}~,
\ee
and if one considers the fermion on its own, one can rescale the frequency according to $\tilde\omega=\omega/\lambda$ 
leading to the IR Lorentz-like fermionic dispersion relation
\be
\tilde\omega^2=\mu_{dyn}^2+p^2+\cdots~,
\ee
where dots represent higher orders in $p^2$ and $\mu_{dyn}\equiv m_{dyn}^3/\lambda$.

\vspace{0.5cm}

\nin{\it Effective light cones}\\
The previous rescaling in the fermion dispersion relation is not valid if one considers that fermions actually 
interact with the scalar, such that both dispersion relations cannot be rescaled independently.
If one takes into account the dynamical generation of the lowest order space derivatives for the scalar, 
rescaling appropriately the frequency leads to the following IR dispersion relations 
\bea\label{disprelY}
\omega^2&=&m_s^2+v_s^2p^2+\cdots~~~~~\mbox{for the scalar}\nn
\omega^2&=&m_f^2+v_f^2p^2+\cdots~~~~~\mbox{for the fermion}~,
\eea
where the finite quantum corrections $v_s$ and $v_f$ naively play the role of effective maximum speeds seen by the scalar and the fermion 
respectively. Because these particles interact, 
it is not possible to find a redefinition of space, time and fields which leads to $v_s=v_f=1$, and both fermions and scalars seem to be 
limited by a maximum speed which is different from the speed of light. But one should take this statement with precaution: the dispersion
relations (\ref{disprelY}) are valid in the IR only, and not for momenta large compared to the the mass of particles. As a consequence, 
$v_s$ and $v_f$ should not be identified with the speed of light.

\section{Gauge theories}\label{gauge}

Detailed discussions and proofs on Lifshitz-type extensions of the Standard Model, as renormalizability, are given in \cite{AnselmiGauge},
where it is also shown that a gauge field has the expected two degrees of freedom, in 3+1 dimensions.

\subsection{Yang-Mills gauge field and detailed balance}

A Lifshitz-type non-Abelian theory is discussed in \cite{Horava}, with the action for $z=2$
\be\label{YM}
S_{YM}=-\frac{1}{4}\mbox{Tr}\int dtd\vec x \left(\frac{2}{e^2}F^{0k}F_{0k}-\frac{1}{g^2}D_iF^{ik}D^jF_{jk} \right)~,
\ee
where $F_{\mu\nu}=\partial_\mu A_\nu-\partial_\nu A_\mu+[A_\mu,A_\nu]$ and $D_i=\partial_i-iA_i$ is the covariant derivative.
Unlike the Abelian case, the Yang-Mills field is self interacting, therefore generating quantum corrections, which 
will dress the couplings $e$ and $g$ differently.\\
One could in principle consider different renormalizable interactions, as
\be
\mbox{Tr}\{F_{ij}F^{jk}F_k^{~i}\}~,~~~\mbox{Tr}\{D_iF_{jk}D^iF^{jk}\}~,
\ee
and in \cite{Horava} is defined the concept of ``detailed balance'', in the following way.
If one calls ``potential'' the term involving space derivatives only, in the action (\ref{YM}), this potential can be obtained from 
the Euclidean Yang Mills action $W$ in $d$ dimensions:
\be
\frac{4}{g^2}D_iF^{ik}D^jF_{jk}=\left( \frac{\delta W}{\delta A_i}\right)\left( \frac{\delta W}{\delta A^i}\right)~,
\ee
where 
\be
W=\frac{1}{2g}\mbox{Tr}\int d\vec x ~F_{ij}F^{ij}~.
\ee
Imposing the detailed balance condition is a way to restrict the number of potential terms that one considers in 
the $d+1$-dimensional Lifshitz model, but should not be seen as a physical restriction.\\
It is then shown in \cite{Horava}, using results from 4-dimensional Yang-Mills theory, that the model (\ref{YM})
in 4+1 dimensions is asymptotically free.

\subsection{Lifshitz-type extensions of $QED$}

A Lifshitz-type extention of $QED$ \cite{AnselmiOther} is, for $d=z=3$,
\bea
{\cal L}&=&-\frac{1}{2}F_{0i}F^{0i}-\frac{1}{4}F_{ij}\left(\Lambda^4\tau_2-\Lambda^2\tau_1\Delta+\tau_0\Delta^2 \right) F^{ij}\\
&&+\ol\psi\left(iD_0\gamma^0+i\Lambda^2b_1\vec D\cdot\vec\gamma-\Lambda b'(\vec D\cdot\vec\gamma)^2
+ib_0(\vec D\cdot\vec\gamma)^3-m \right) \psi~,\nonumber
\eea
where $\Lambda$ is an energy scale, $D_\mu$ is the covariant derivative 
and the parameters $\tau_0,\tau_1,\tau_2,b_0,b_1,b'$ are dimensionless. 
Note that, due to gauge invariance, higher order space derivatives need to be covariant, therefore involving new vertices with 2 or 3 photons and 2
fermions.
The gauge field has the following dispersion relation, after rescaling of the frequency,
\be
\omega^2=p^2\left(\tau_2+\tau_1\frac{p^2}{\Lambda^2}+\tau_0\frac{p^4}{\Lambda^4} \right) 
\ee
The hierarchy problem is discussed in \cite{AnselmiOther} and it is shown that hierarchy is smoothend by the anisotropic feature 
of this mode.
This problem of mass hierarchy, with flavour mixing and with higher dimensions, is also discussed in \cite{KanetaKawamura}, and 
proton stability can be obtained from anisotropic scaling \cite{Kawamura}, which implies different relations between unification scales.

\vspace{0.5cm}

\nin{\it UV completion}\\
Another Lifshitz-type extention of $QED$ is proposed in \cite{IengoSerone2}, for $d=4$ and $z=2$, where the authors show that the corresponding 
model exhibits UV completion: the gauge coupling is finite at high energies, to all orders in perturbation theory. In the IR,
the model recovers usual 5-dimensional $QED$, with the 4th space dimension compactified on a circle. Nevertheless, the problem of 
mismatch between effective light cones for the photons and fermions is still present.

\subsection{An alternative Lorentz-violating $QED$}\label{alternative}

To finish this review we discuss some features of an alternative Lorentz-violating Abelian gauge theory in 3+1 dimensions \cite{AlexMavroVergou},
which also contains higher-order space derivatives, but is not Lifshitz-type in the sense that space ant time have same mass dimension.\\
The massless fermion Lagrangian is
\be\label{LVQED}
{\cal L}=-\frac{1}{4}F^{\mu\nu}\left(1-\frac{\Delta}{M^2}\right)F_{\mu\nu}+i\ol\psi\br D\psi~,
\ee
where the Lorentz-violating term has two roles: introduce a mass scale, necessary to generate a fermion mass, 
and lead to finite gap equation for the fermion mass. Such higher order space derivatives can arise in a 
low-energy effective action of open strings propagating in D0-particle stochastic space time foam backgrounds \cite{D0}. 
We stress here that $M$ regularizes only loops with an internal photon line, and that
another regularization is necessary to deal with fermion loops. 
Also, the Lorentz violating modifications proposed in the Lagrangian (\ref{LVQED}) do not 
alter the photon dispersion relation, which remains relativistic.\\
It is shown in \cite{AlexMavroVergou} that fermion dynamical mass generation occurs for the model 
(\ref{LVQED}) and, in the Feynman gauge, gives
\be\label{mdyn}
m_{dyn}=M\exp\left(-\frac{\pi}{2\alpha}\right)~,
\ee
where $\alpha\equiv e^2/4\pi$ is the fine structure constant. The gauge dependence of this result can be
circumvented in the following way. It has been argued, using the pinch technique \cite{pinch}, that the calculation of 
physical quantities in a gauge theory can be done with the Feynman gauge. The reason for this is
the cancellation of longitudinal contributions to the gauge propagator during the 
perturbative expansion. As a result of these cancellations, the calculation of the physical quantity leads to the
result that would be obtained in the Feynman gauge.\\ 
One needs to considers $M$ as a regulator (not for the expression (\ref{mdyn}) though, where $M$ could be identified with the 
Plank mass), 
in order not to break down the loop structure of the theory. Indeed, the divergences generated by this regulator have the form
$\alpha\ln(m_{dyn}/M)$, which is of order zero loop. These terms must therefore be included in the counter terms of the 
theory. It is then shown in \cite{AlexMavroVergou} that quantum corrections, 
which are not the same for space and time derivatives of the fermion, 
lead to an effective maximum speed for fermions
\be 
v^2=1-\frac{2\alpha}{\pi}\left(\frac{34}{9}-\frac{16}{3}\ln2\right) +{\cal O}(\alpha^2)~< ~1~.
\ee
The speed of the gauge field remains $c=1$, to all orders in $\hbar$.

Extensions of the model (\ref{LVQED}) are then discussed \cite{AlexMavroVergou}, where a possible dynamical Higgs mechanism is proposed.
Indeed, it is known that flavour symmetry breaking implies gauge symmetry breaking \cite{dynHiggs}, through a mechanism similar to the
Schwinger mechanism in 1+1-dimensional $QED$. In \cite{AlexMavroVergou}, flavour dynamical symmetry breaking is induced by
Lorentz-violating higher order space derivatives for the vector boson.

\section{Conclusion}

Lorentz symmetry violation has apparently been observed with the superluminal propagation of neutrinos \cite{Opera}, 
and other signals might be detected at the LHC, and can lead to a rich phenomenology.
Such effects can be taken into account by Lifshitz-type theories, which can therefore play the role of bridge between 
experimental data and fundamental origins, like Quantum Gravity.\\
Fundamental properties of Lifshitz-type theories have been described in this short review, which can be generalized to other models. 
Some topics were not discussed here, as: {\it (i)} stochastic quantization and the analogy with Langevin equation 
in 2+1 dimensions, for $z=2$ \cite{OrlandoReffert}; {\it (ii)} Lifshitz-Schwinger model, with fermionization 
and bosonization of 1+1 dimensional gauge models \cite{SonKim}; {\it (iii)} gauged scalar theory in 1+1 dimensions \cite{EuneKimSon2};
{\it (iv)} monopoles and confinement for Lifshitz-type $QED$ in 2+1 dimensions \cite{DasMurthy2+1}.\\
Because dynamical mass generation occurs naturally in the framework of Lifshitz-type theories, there is the possibility to
build an alternative to the Standard Model, without Higgs particle. In this context, Lorentz symmetry violation can trigger 
the dynamical breaking of symmetries, and generate a dynamical Higgs mechanism. The corresponding phenomenology 
remains to be studied thouroughly.\\
We finish with a problem related to Horava-Lifshitz Gravity \cite{HoravaGravity}, and which does not occur in 
flat space time Lifshitz theories. A known problem of Horava-Lifshitz Gravity
is the existence of an additional scalar graviton degree of freedom, as the consequence of the breaking of full diffeomorphism 
in space time \cite{diffeo}. A solution to this problem was suggested in \cite{covariantHL}, where auxiliary fields are introduced,
giving rise to an additional equation of motion, therefore eliminating the unwanted scalar graviton. Several articles were published 
on this extension of Horava-Lifshitz Gravity \cite{several}, but the question of how to retrieve uniquely General Relativity in the IR
is not solved yet.

\end{document}